\documentclass[prb,reprint,superscriptaddress,showpacs,amsmath,amssymb,floatfix,
]{revtex4-1}
\usepackage{dcolumn}
\usepackage{bm}
\usepackage{turnstile}
\usepackage{graphicx}
\usepackage{amsmath,amssymb}
\usepackage{url}
\usepackage{multirow}
\usepackage{float}
\usepackage[toc,page]{appendix}
\usepackage{color}
\usepackage{ulem}

\DeclareMathOperator{\tr}{\mathop{\mathrm{Tr}}}

\DeclareMathOperator{\const}{\mathop{\mathrm{const}}}

\begin{document}
\title{Hybrid helical state and superconducting diode effect in S/F/TI heterostructures}
\author{T.~Karabassov}\email{tkarabasov@hse.ru}
	\affiliation{HSE University, 101000 Moscow, Russia}
\author{I.~V.~Bobkova}
    \affiliation{Institute of Solid State Physics, Chernogolovka, Moscow reg., 142432 Russia}
    \affiliation{Moscow Institute of Physics and Technology, Dolgoprudny, 141700 Russia}
    \affiliation{HSE University, 101000 Moscow, Russia}
\author{A.~A.~Golubov}
	\affiliation{Faculty of Science and Technology and MESA$^+$ Institute for Nanotechnology,
		University of Twente, 7500 AE Enschede, The Netherlands}
\author{A.~S.~Vasenko}
    \affiliation{HSE University, 101000 Moscow, Russia}
    \affiliation{I.E. Tamm Department of Theoretical Physics, P.N. Lebedev Physical Institute, Russian Academy of Sciences, 119991 Moscow, Russia}

\begin{abstract}
It is well-known that the ground state of homogeneous superconducting systems with spin-orbit coupling (SOC) in the presence of the Zeeman field is the so-called helical state, which is characterized by the phase modulation of the order parameter, but zero supercurrent density. In this work we investigate the realization of the helical state in  a hybrid system with spatially separated superconductivity and exchange field by considering superconductor/ferromagnet (S/F) bilayer on top of a 3D topological insulator. This system is characterized by strong spin-momentum locking and, consequently, provides the most favorable conditions for the helical state generation. The analysis is based on the microscopic theory in terms of the quasiclassical Green’s functions. We demonstrate that in the bilayer the helical state  survives if the exchange field has non-zero component perpendicular to the S/F interface even in spite of the fact that the superconducting order parameter and the exchange field are spatially separated. At the same time, in this spatially inhomogeneous situation the helical state is accompanied by the spontaneous currents distributed over the bilayer in such a way  that the net current vanishes. Further, we show that this hybrid helical state gives rise to nonreciprocity in the system. We demonstrate the realization of the nonreciprocity in the form of the superconducting diode effect  and investigate its dependence on the parameters of bilayer.

\end{abstract}
\maketitle

\section{Introduction}
 
 The helical state was originally predicted for two-dimensional systems with spin-orbit coupling (SOC) under the applied parallel magnetic field  \cite{Edelstein1989,Barzykin2002,Dimitrova2007,Samokhin2004,Kaur2005,Houzet2015}. Its physical origin can be explained as follows. The SOC produces the spin-momentum locking term $\propto \bm n \cdot (\bm \sigma \times \bm p)$ in the hamiltonian, where $\bm n$ is the unit vector  perpendicular to the plane of the system, $\bm p$ is the electron momentum and $\bm \sigma$ is its spin. The applied field makes spin-down state energetically more favorable. Due to the spin-momentum locking it results in the fact that one of the mutually opposite momentum directions along the axis perpendicular to the Zeeman field is more favorable. That should lead to the appearance of the spontaneous current. However, the superconductor develops a phase gradient, which exactly compensates the spontaneous current. The resulting phase-inhomogeneous zero-current state is the true ground state of the system. This helical state is a kind of inverse magnetoelectric effect specific for superconductors. This state looks similar to another well-known inhomogeneous superconducting state, FFLO state \cite{Fulde1964,Larkin1965,Mironov2012,Mironov2018}. However, the crucial difference between them is that in the helical state the direction of the phase modulation is strictly determined by the direction of the applied field, while in the FFLO state the direction of the modulation is mainly determined by the crystal structure. The same physics can be also expected if the Zeeman field is provided not by the applied magnetic field, but by the intrinsic exchange field. In this case the helical state provides a direct coupling between the condensate phase and the magnetization, which opens great perspectives for superconducting spintronics. 

The situation when the exchange field, superconductivity and strong SOC coexist intrinsically is rare and largely unexplored from the point of view of magnetoelectrics. At the same time the interplay of superconductivity and magnetism is actively studied in superconductor/ferromagnet (S/F) hybrids \cite{Buzdin2005,Bergeret2005,Eschrig2015,Linder2015}, where the order parameter and the exchange field are spatially separated. In the presence of spin-momentum locking a plethora of extremely interesting magnetoelectric effects in the form of spontaneous currents have been reported in the literature for such a situation \cite{Bobkova2004,Mironov2017,Pershoguba2015,Malshukov2020,Malshukov2020_2,Malshukov2020_3,Alidoust2017}. Josephson junctions deserve special mention, where the magnetoelectric effect manifests itself in the form of the anomalous ground state phase shift \cite{Krive2004,Nesterov2016,Dolcini2015,Reynoso2008,Buzdin2008,Zazunov2009,Brunetti2013,Yokoyama2014,Bergeret2015,Campagnano2015,Konschelle2015,Kuzmanovski2016,Malshukov2010,Rabinovich2019,Alidoust2020}. 

Here we consider finite-width S/F bilayer on top of a three-dimensional topological insulator (3D TI). 3D TI is chosen because its conductive surface state exhibits full spin-momentum locking: an electron spin always makes a right angle with its momentum \cite{Burkov2010,Culcer2010,Yazyev2010,Li2014}.   It has been already predicted that for this system presence of the helical magnetization in the F layer leads to the nonmonotonic dependence of the critical temperature on the F layer width \cite{Karabassov2021}. Here we consider another important manifestation of the interplay between the spin-momentum locking and the magnetization in this system. It is found that although the exchange field and superconducting order parameter are spatially separated, the latter develops a spontaneous phase gradient, that is the finite-momentum helical state is realized. At the same time it is accompanied by the spontaneous currents, inhomogeneously distributed over the bilayer in such a way that the net current vanishes. Such a hybrid state only takes place when the exchange field has a component perpendicular to the S/F interface. Otherwise, the bilayer is in the conventional homogeneous state. 

Further we demonstrate that this hybrid state is intrinsically nonreciprocal, that is the bilayer possesses different critical currents in opposite directions. In the literature this phenomenon is also referred to as the “superconducting diode effect (SDE)”. A superconductor exhibiting such a polarity-dependent critical current is of interest both from fundamental and applied points of view. It can offer a perfect dissipationless transmission along one direction while manifesting a large resistance along the opposite.  It represents the superconducting limit of the magnetochiral anisotropy (MCA) \cite{Rikken2001,Krstic2002,Pop2014,Rikken2005,Ideue2017,Wakatsuki2018,Hoshino2018,Wakatsuki2017,Qin2017,Yasuda2019,Itahashi2020,Lin_arxiv}. The effects are being actively studied during the last few years. The superconducting diode effect has been predicted for homogeneous materials with SOC and finite-momentum helical ground state \cite{Daido2022,He_arxiv,Yuan_arxiv,Scammell_arxiv,Ilic_arxiv,Legg2022_arxiv}, S/F bilayers with interface spin-orbit coupling \cite{Devizorova2021} and for Josephson junctions \cite{Yokoyama2014,Kopasov2021,Davydova_arxiv,Halterman2022,Alidoust2021,Tanaka2022_arxiv,Golod2022_arxiv,Kokkeler2022}. It has been also observed in superconducting films, layered systems without a centre of inversion\cite{Ando2020,Bauriedl_arxiv,Shin_arxiv,Hou_arxiv,Hideki_arxiv} and Josephson junctions \cite{Bocquillon2017,Baumgartner2022,Wu2022,Pal_arxiv,Baumgartner_arxiv,Zhang_arxiv,Hu2007,Chen2018}. Here we investigate it in the topological insulator based systems. Our consideration is based on the microscopic quasiclassical theory of superconductivity in terms of the Usadel equations. 

The paper is organized as follows. In Sec.~\ref{Model} we formulate basic theory in the framework of the quasiclassical Usadel formalism. In Sec.~\ref{q_state} the hybrid helical state is investigated and in Sec.~\ref{I_q} we show the presence of the current nonreciprocity and  present the results of the SDE calculation in the system. Finally, we summarize the key points of the research in Sec.~\ref{Fin}.

\section{Model}\label{Model}
\begin{figure*}[t]
	\includegraphics[width=1.8\columnwidth,keepaspectratio]{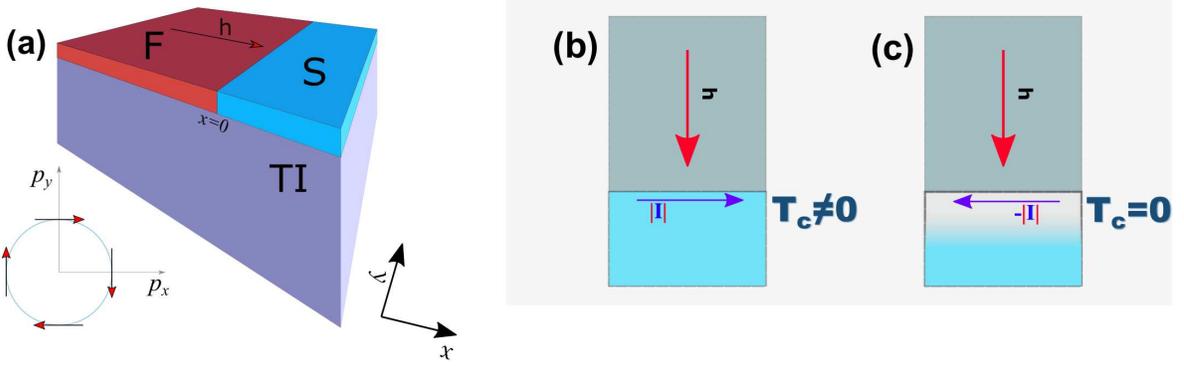}
	\caption{ (a)  Schematic geometry of the S/F bilayer on top of the 3D TI. Left bottom corner: Fermi-surface of the TI surface states. The quasiparticle spin $\bm S$ is locked at the right angle to its momentum $\bm p$.   (b)-(c) Illustration of the superconducting diode effect. Applying external supercurrent along the interface in one direction keeps the non-zero critical temperature (b), while reversing the current may completely destroy the superconducting state (c). 
	}
	\label{model_pic}
\end{figure*}
In the present work we consider  an S/F bilayer on top of a 3D TI surface. It is sketched in Fig.~\ref{model_pic} (a). The F layer is assumed to be a ferromagnetic insulator and it induces an exchange field in the conductive surface states of the 3D TI underneath via the proximity effect. Experimental realization of such a  proximity-induced exchange field has been reported \cite{Jiang2014,Wei2013,Jiang2015,Jiang2016}. Similarly, the superconductor provides proximity-induced superconductivity in the conductive surface states of the 3D TI underneath \cite{Fu2008}. The resulting hamiltonian of the 3D TI conductive surface layer takes the form:
 \begin{eqnarray}
H=\int d^2 r \Biggl\{\Psi^\dagger (\bm r)\bigl[-i\alpha(\bm \nabla_{\bm r}\times \hat z)\bm \sigma - \mu +  V(\bm r)- \nonumber \\
\bm h \bm \sigma \bigr]\Psi(\bm r)+\Delta(\bm r)\Psi^\dagger_\uparrow (\bm r) \Psi^\dagger_\downarrow (\bm r) + \Delta^*(\bm r)\Psi_\downarrow (\bm r) \Psi_\uparrow (\bm r) \Biggr\},~~~~
\label{ham}
\end{eqnarray}
where $\Psi^\dagger(\bm r)=(\Psi^\dagger_\uparrow(\bm r),\Psi^\dagger_\downarrow(\bm r))$ is the creation operator of an electron at the 3D TI surface, $\hat z$ is the unit vector normal to the surface of TI, $\alpha$ is the Fermi velocity of electrons at the 3D TI surface and $\mu$ is the chemical potential. $\bm \sigma = (\sigma_x, \sigma_y, \sigma_z)$ is a vector of Pauli matrices in spin space and $\bm h = (h_x, h_y, 0)$ is an in-plane exchange field, which is assumed to be nonzero only at $x<0$. The superconducting order parameter $\Delta $ is nonzero only  at $x>0$. Therefore, effectively the TI surface states are divided into two parts: one of them at $x<0$ possesses $h \neq 0$ and can be called "ferromagnetic", while the other part corresponding to $x>0$ with $\Delta \neq 0$ can be called "superconducting". Below we will use subscripts $f$ and $s$ to denote quantities, related to the appropriate parts of the TI surface. The potential term $V(\bm r)$ includes the nonmagnetic impurity scattering potential $V_{imp}=\sum \limits_{\bm r_i}V_i \delta(\bm r - \bm r_i)$, which is of a Gaussian form $\langle V(\bm r)V(\bm r')\rangle = (1/\pi \nu \tau)\delta(\bm r - \bm r')$ with $\nu=\mu/(2\pi \alpha^2)$, and also possible interface potential $V_{int}(\bm r) = V\delta(x)$.

We consider the situation when $\mu$ is large. In this case the Fermi surface  is represented by a single helical band, where the electrons manifest the property of the full spin-momentum locking, see Fig.~\ref{model_pic}(a).  Due to the large value of $\mu$ the quasiclassical approximation is the well-suited framework to describe the system.  Here we assume the diffusive limit, i. e. when the elastic scattering length is much smaller than the superconducting coherence length ($l \ll \xi_s$). In this situation the system should be described by the diffusion-type Usadel equations for the quasiclassical Green's function, which have been derived in Refs.~\onlinecite{Zyuzin2016} and \onlinecite{Bobkova2016}. We begin by considering the linearized with respect to the anomalous Green's function Usadel equations in Matsubara representation. The linearization works well near the critical temperature of the bilayer, when the superconducting order parameter is small. Therefore, this framework is enough to calculate the critical temperature and to investigate the superconducting state near the critical temperature. Further we turn to the nonlinear Usadel equations in order to calculate the supercurrent and to study the SDE. In principle, the anomalous Green's function is a $2 \times 2$ matrix in spin space. However, its spin structure is determined by the projection onto the conduction band of the TI surface states and, therefore, one can write:
\begin{equation}
\hat f_{s,f}(\bm n_F, \bm r, \varepsilon)= f_{s,f}(\bm r, \varepsilon)\frac{(1+\bm n_\perp \bm \sigma)}{2},
\label{spin_structure}
\end{equation}
where $\hat f_{s(f)}$ is the anomalous Green's function in the superconducting (ferromagnetic) part of the 3D TI layer. $\bm n_F=\bm p_F/p_F=(n_{F,x},n_{F,y},0)$ is a unit vector directed along the quasiparticle trajectory and $\bm n_\perp=(n_{F,y},-n_{F,x},0)$ is a unit vector perpendicular to the quasiparticle trajectory and directed along the quasiparticle spin, which is locked to the quasiparticle momentum. $f_{s,f}$ is the {\it spinless} amplitude of the Green's function, which describes mixed singlet-triplet correlations in the system and in the diffusive limit is isotropic in the momentum space. 

Our first goal is to calculate the critical temperature $T_c$ of the structure. We assume that the superconducting layer is ultra-thin along the $z$-direction. In the framework of our model it is considered as two-dimentional and is described by Hamiltonian (\ref{ham}). Strictly speaking, the S film as a whole is not described by Hamiltonian (1), but, nevertheless, it has a strong spin-orbit coupling induced by proximity to the 3D TI. It results in qualitatively the same structure of the Green's function, but requires much more sophisticated modelling. In order to focus on the main physical properties of the mixed helical state and the nonreciprocity, we work in the framework of the minimal model.  In the superconducting  part of the TI conductive surface (S) ($0 < x <d_s$) the linearized Usadel equation for the spinless amplitude $f_s$ reads\cite{Belzig1999,Usadel,Zyuzin2016,Bobkova2017}
\begin{equation}\label{Usadel_S}
\xi_s^2 \pi T_{cs} \left(\partial_x^2 + \partial_y^2 \right) f_s - |\omega_n| f_s + \Delta(\textbf{r}) = 0.
\end{equation}
Units with $\hbar= k_B= 1$ are used.
In the ferromagnetic part of the TI conductive surface layer (F) the Usadel equation takes the form \cite{Zyuzin2016},
\begin{equation}\label{Usadel_TI}
\left(\partial_x -\frac{2i}{\alpha} h_y\right)^2 f_f + \left(\partial_y +\frac{2i}{\alpha} h_x\right)^2 f_f = \frac{|{\omega_n}|}{\xi_f^2 \pi T_{cs}} f_f.
\end{equation}
In Eqs.~(\ref{Usadel_S}) and (\ref{Usadel_TI}) $\xi_{s(f)} = \sqrt{D_{s(f)}/ 2 \pi T_{cs}}$, where $D_{s(f)}$ is the diffusion constant in S(F) region, which, in principle, can be different due to the coverage of the TI by different materials in those parts, and $T_{cs}$ is the critical temperature of the bulk superconductor. In order to account for the helical state we consider the pair potential to be of the form,
\begin{equation} \label{Delta_q}
\Delta (x,y)=\Delta (x) e^{i q y}.
\end{equation}
Then the anomalous Green's function in the S  part of the TI have to manifest the same dependence on $y$-coordinate:
\begin{equation}
f_s (x,y)=f_s (x) e^{i q y}.
\end{equation}
The Usadel equation in the S  part then becomes one-dimensional and takes the form:
\begin{equation}\label{Usadel_S_q}
\xi_s^2 \pi T_{cs} \left(\partial_x^2 -q^2\right) f_s - |\omega_n| f_s + \Delta = 0.
\end{equation}
In the ferromagnetic region of the TI we assume only  the nonzero $h_x$ component of the field and utilize the same ansatz as in the S part, i. e. $f_f (x,y)=f_f (x) e^{i q y}$ as it is dictated by the boundary conditions,
\begin{equation} \label{Usadel_TIq}
\partial_x^2 f_f= \left[\frac{|{\omega_n}|}{\xi_f^2 \pi T_{cs}} +  \left( q + \frac{2 h_x}{\alpha}\right)^2\right]f_f.
\end{equation}
Inclusion of the magnetization component $h_y$ produces no quantitative effect neither on the supercurrent in $y$ direction of the bilayer nor on the critical temperature in the S part. It only enters the solution $f_f$ as a phase factor $exp\left(2 i h_y x/\alpha\right)$ \cite{Zyuzin2016,Karabassov2021}. Thus we do not take it into consideration in our model and define $h_x=h$.

The self-consistency equation in the S part of the system can be written as
\begin{align}
\Delta \ln \frac{T_{cs}}{T} = \pi T \sum_{\omega_n} \left ( \frac{\Delta}{|\omega_n|} - f_s \right ).
\end{align}
We also need to supplement the equations above with proper boundary conditions\cite{KL,Zyuzin2016} at $x=0$. Due to the fact that the spin structure of the Green's functions at the both sides of the interface is the same, the boundary conditions can be written in terms of the spinless Green's functions and take the form
\begin{equation}\label{KL1}
\gamma_B \xi_f \frac{\partial  f_f(0)}{\partial x} = f_s(0) - f_f(0) ,
\end{equation} 
\begin{equation}\label{KL2}
\gamma \xi_f \frac{\partial  f_f(0)}{\partial x}= \xi_s \frac{\partial f_s(0)}{\partial x}.
\end{equation} 
The parameter $\gamma_B=R_b \sigma_f / \xi_f$ is the transparency parameter which is the ratio of resistance per unit area of the effective S/F interface at $x=0$ to the resistivity of the  ferromagnetic part of the TI surface and describes the effect of the interface barrier \cite{KL, gammab1, gammab2}. In Eq.\eqref{KL2} the dimensionless parameter $\gamma = \xi_s \sigma_f / \xi_f \sigma_s$ determines the strength of suppression of superconductivity in the S near the S/F interface compared to the bulk (inverse proximity effect). 
No suppression occurs for $\gamma = 0$, while strong suppression takes place for $\gamma \gg 1$. Here $\sigma_{s(f)}$ is the normal-state conductivity of the S(F) parts of the TI surface.
These boundary conditions should also be supplemented with vacuum conditions at the edges ($x=-d_f$ and $x=+d_s$),
\begin{align}\label{Vacuum}
\frac{\partial f_s(d_s)}{\partial x}=0, \quad
\frac{\partial  f_f(-d_f)}{\partial x}  =0.
\end{align}
The solution of Eq.~\eqref{Usadel_TIq} can be found in the form
\begin{equation}\label{f_T}
f_f= C(\omega_n) \cosh k_{q} \left( x + d_f \right),
\end{equation}
where
\begin{align}\label{kq}
k_{q} = \sqrt{\frac{|{\omega_n}|}{\xi_f^2 \pi T_{cs}} + \left( q + \frac{2 h}{\alpha}\right)^2}.
\end{align}
Here $C(\omega_n)$ is to be found from the boundary conditions. Eq.~(\ref{f_T}) automatically satisfies the vacuum boundary condition \eqref{Vacuum} at $x=-d_f$.
Using boundary conditions \eqref{KL1}-\eqref{KL2} we can write the problem in a closed form with respect to the Green function $f_s$. At $x=0$ the boundary conditions can be written as:
\begin{equation}\label{boundary}
\xi_s \frac{\partial f_s(0)}{\partial x} = \frac{\gamma}{\gamma_B + A_{qT} (\omega_n)} f_s(0),
\end{equation}
where,
\begin{align*}
A_{qT}(\omega_n)= \frac{1}{k_{q} \xi_f} \coth{ k_{q} d_f}.
\end{align*}
Then, we rewrite the Usadel equation in the S part of the TI surface in terms of $f_s^+$ and $f_s^-$, where we define even and odd parts of the anomalous Green's function $f^\pm = f(\omega_n) \pm f(-\omega_n)$. According to the Usadel equation \eqref{Usadel_S}, there is a symmetry relation $f(-\omega_n) = f^*(\omega_n)$ which implies that
$f^+$ is a real while $f^-$ is a purely imaginary function.
In general the boundary condition \eqref{boundary} can be complex. But in the considered system $A_{qT}$ is real. Hence the condition \eqref{boundary} coincides with its real-valued form,
\begin{equation}\label{B1a}
\xi_s \frac{\partial f^+_s(0)}{\partial x} = W^{q}(\omega_n) f^+_s(0),
\end{equation}
where we used the notation,
\begin{align}\label{W_om}
W^{q}(\omega_n) &= \frac{\gamma}{\gamma_B + A_{qT} (\omega_n)}.
\end{align}
In the same way we rewrite the self-consistency equation for $\Delta$ in terms of symmetric function $f_s^+$ considering only positive Matsubara frequencies,
\begin{equation}\label{Delta+}
\Delta \ln \frac{T_{cs}}{T} = \pi T \sum_{\omega_n > 0} \left ( \frac{2\Delta} {\omega_n} - f_s^{+} \right),
\end{equation}
as well as the Usadel equation in the superconducting part,
\begin{equation}\label{finUsadelS}
\xi_s^2 \left(\frac{\partial^2 f^{+}_s}{\partial x^2} - \kappa_{qs}^2 f^{+}_s\right) + \frac{2 \Delta}{\pi T_{cs}} = 0.
\end{equation}
In the framework of the so-called single-mode approximation the solution in S is introduced in the form\cite{Fominov2002},
\begin{equation}\label{Fssingle}
f_s^+(x,\omega_n)=f(\omega_n) \cos\left(\Omega\frac{x-d_s}{\xi_s}\right),
\end{equation}
\begin{equation} \label{Dsingle}
\Delta(x)=\delta \cos \left(\Omega\frac{x-d_s}{\xi_s}\right).
\end{equation}
The solution presented above automatically satisfies boundary condition \eqref{B1a} at $x=d_s$.
Substituting expressions \eqref{Fssingle} and \eqref{Dsingle} into the Usadel equation for $f_s^+$ \eqref{finUsadelS} yields
\begin{align}\label{f_om}
f(\omega_n)=\frac{2 \delta}{\omega_n + \Omega^2 \pi T_{cs} +q^2 \xi_s^2 \pi T_{cs}}.
\end{align}
 In the following section we calculate the critical temperature $T_c$ using the equations above. Exact results for the anomalous Green's function and the critical temperature can be obtained in the framework of the more sophisticated multi-mode approach. However, for the case under consideration the multi-mode approach gives only quantitative corrections to the results, as it is shown in the Appendix.

\section{Hybrid superconducting helical state}

\label{q_state}

To calculate the critical temperature  we use Eqs.~ \eqref{B1a}-\eqref{finUsadelS}, together with the vacuum boundary conditions \eqref{Vacuum} for the anomalous Green's function $f^{+}_s$. Further we assume $\xi_s=\xi_f=\xi$ in our calculations for clarity and simplicity of the results. 
Using the single-mode approximation \eqref{Fssingle}-\eqref{Dsingle} it is possible to rewrite the self-consistency equation in the following form,
\begin{equation}\label{last1}
\ln \frac{T_{cs}}{T_c} = \psi \left ( \frac{1}{2} + \frac{\Omega^2 + q^2 \xi^2}{2}\frac{T_{cs}}{T_c} \right ) - \psi \left ( \frac{1}{2} \right ).
\end{equation}
Boundary condition \eqref{B1a} at $x=0$ yields the following equation for $\Omega$,
\begin{equation}\label{last2}
\Omega \tan \left ( \Omega \frac{d_s}{\xi} \right ) = W^{q}(\omega_n).
\end{equation}
The finite momentum  of the pair amplitude $q=q_s$, which is chosen by the system, is determined by the condition,
\begin{equation}\label{cond}
q_s = q\left(\max \left[T_c(q)\right]\right),
\end{equation}
which means that the state with $q_s$ is the most energetically favorable. 
\begin{figure}[t]
	\includegraphics[width=\columnwidth,keepaspectratio]{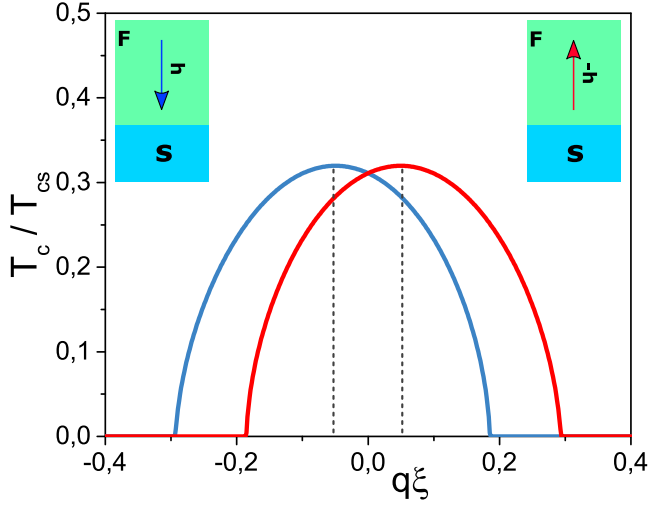}
	\caption{$T_c(q)$ dependencies for $  \xi h/ \alpha=0.95$, $d_f=1.0 \xi$, $d_s=1.2 \xi$. The parameters of the S/F interface: $\gamma=0.2$, $\gamma_B=0.1$. }
	\label{fig1}
\end{figure}
We can expect that in the absence of magnetization $h$ the equilibrium value of $q_s$ is zero. At nonzero $h$ the equilibrium pair momentum $q_s$ is finite. The dependence of the critical temperature on the pair momentum $q$ is demonstrated in Fig.~\ref{fig1} for two opposite  values of the magnetization strength $h$. According to Eq.\eqref{cond} the most favorable superconducting state corresponds to $ q_s \xi \approx  \pm 0.05$ for $\xi h/ \alpha = \mp 0.95$. This observation indicates that conventional superconducting state undergoes a qualitative change. The superconducting gap $\Delta$ is now modulated with a phase factor $\exp(i q_s y)$ generating corresponding phase gradient along the S/F interface. In fact, as we will show below the supercurrent caused by $q_s$ exactly compensates the supercurrent flowing on the TI surface in the opposite direction.

\begin{figure}[t]
	\includegraphics[width=\columnwidth,keepaspectratio]{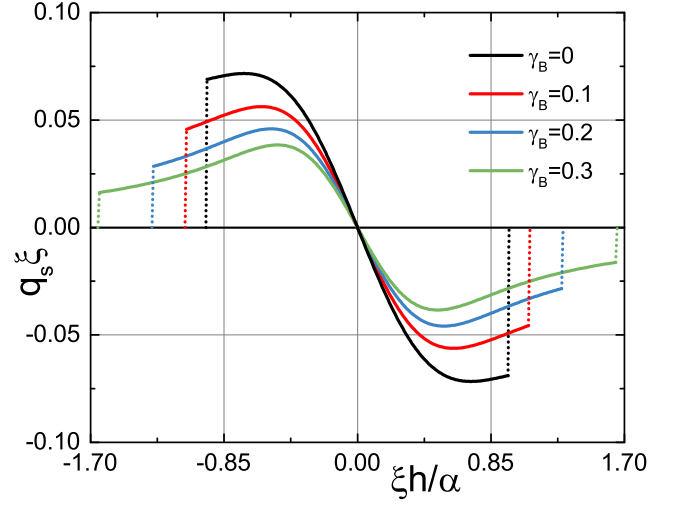}
	\caption{ The ground state pair momentum $q_s(h)$ for different values of transparency parameter $\gamma_B$. The other parameters are the same as in Fig.~\ref{fig1} }
	\label{fig2}
\end{figure}

The dependence of $q_s$ on $h$ is demonstrated in Fig.~\ref{fig2}. We plot the curves for different values of the interface transparencies $\gamma_B$. From the figure we can see that for the transparent interface ($\gamma_B=0$) the pair momentum $q_s$ is the most pronounced.  Physically it just reflects the necessity of the proximity to the ferromagnetic layer to produce the hybrid helical state. Abrupt drop to zero of the parameter $q_s$ reflects the transition from superconducting to normal state.
%

Under the assumption $d_s \ll \xi_s$ the solution in the system can be analyzed analytically. In this case we can assume the gap $\Delta (x)$ to be spatially constant $\Delta=\const$. Then the solution in the superconducting region takes the form
\begin{align}\label{thins_sol}
f_s^+&=\frac{2 \Delta}{\omega_n + q^2 \xi^2 \pi T_{cs}} + A(\omega_n) \cosh \left( \kappa_{qs} \left[x- d_s\right] \right), \\
A(\omega_n)&= - \frac{2 \Delta}{\cosh \left( \kappa_{qs} d_s \right) \left[\omega_n + q^2 \xi^2 \pi T_{cs}\right]} \frac{W^q(\omega_n)}{W^q(\omega_n) + A_{qs}} \nonumber, \\
A_{qs} &= \kappa_{qs} \xi \tanh \kappa_{qs} d_s, \quad 
\kappa_{qs} =\sqrt{q^2 +\frac{|{\omega_n}|}{\xi^2 \pi T_{cs}}}. \nonumber
\end{align}
Here, to find the coefficient $A(\omega_n)$ the boundary condition \eqref{B1a} has been utilized.
On the other hand, the solution in TI layer is,
\begin{align}\label{F_sol_an}
    f_f= \frac{f_s(0)}{\gamma_B + A_{qT}} \frac{\cosh(k_q [x+d_f])}{k_q \xi \sinh(k_q d_f)}.
\end{align}
In the limit  $d_f\ll \xi$ and $\gamma_B=0$ we can derive analytical result for the critical temperature at the interface. From the analytical solution in the S part of the TI layer we find,
\begin{align}
    f_s^+(0)=\frac{2 \Delta}{\omega_n + (q \xi)^2 \pi T_{cs}} \frac{A_{qs}}{A_{qs}+W^q(\omega_n)}.
\end{align}
Substitution of this expression into the self-consistency equation yields,
\begin{align}\label{Tc_a}
    \ln{\frac{T_{cs}}{T_c}} = \frac{d_s }{\gamma d_f +d_s} \psi\left( \frac{1}{2} + \frac{\gamma d_f (q \xi + H)^2 + q^2 d_s \xi }{2 (\gamma d_f + d_s) T_c / T_{cs} \xi} \right) - \nonumber \\ -\psi\left( \frac{1}{2} \right),
\end{align}
where $H=2 \xi h /\alpha $. It is worth considering Eq.~\eqref{Tc_a} in the limiting case of small $q$. Expanding the equation up to the second order in $q$ we obtain, 
\begin{align}
    \ln{\frac{T_{cs}}{T_c}}= \left( \frac{d_s }{\gamma d_f +d_s} - 1\right)  \psi\left( \frac{1}{2} \right) + \nonumber \\
    + \frac{ \pi^2 \left( \gamma (d_s d_f/\xi^2) \left( q \xi + H\right)^2 + (q d_s)^2\right)}{4 T_c \left( \gamma d_f + d_s\right)^2}.
\end{align}
From this expression we can easily derive important analytical result for the finite momentum $q_s$ of the pair potential $\Delta$. Utilizing the condition for finding extrema of $T_c(q)$ we get,
\begin{align}\label{qs_a}
    q_s = - \frac{\gamma H d_f }{d_s+ \gamma d_f }.
\end{align}
Under the assumption of small $\gamma$, we can approximate the hybrid helical state momentum as $q_s \propto - \gamma H d_f /d_s$. We clearly see that $q_s$ depends on the dimensionless product  $H d_f /d_s$. As we will show below the superconducting diode effect is also controlled by the same parameter.

In contrast to the well-known helical state in homogeneous systems in the presence of the SOC and Zeeman field, where the finite-momentum equilibrium state corresponds to zero supercurrent density, here  the finite-momentum Cooper pairs coexist with nonzero supercurrent density in the ground state of the system. Below we calculate the spatial distribution of the supercurrent for a given $q$. 

In order to calculate the supercurrent we consider the nonlinear Usadel equation, which is of the form\cite{Zyuzin2016,Bobkova2017}
\begin{equation}\label{Usadel_general}
D \hat{\nabla}\left(\hat{g} \hat{\nabla} \hat{g} \right)= \left[\omega_n \tau_z + i \hat{\Delta}, \hat{g}\right].
\end{equation}
Here $D$ is the diffusion constant, $\tau_z$ is the Pauli matrix in the particle-hole space, $\hat{\nabla} X = \nabla X + i \left(h_x \hat{e}_y - h_y \hat{e}_x\right) \left[\tau_z, \hat{g}\right]/\alpha$. The gap matrix $\hat{\Delta}$ is defined as $\hat{\Delta}= \hat{U} i \tau_x \Delta (x) \hat{U}^\dagger$, where $\Delta(x)$ is a real function and transformation matrix $\hat{U}= \exp \left( i q y \tau_z/2 \right)$ . The finite center of mass momentum $q$ takes into account the helical state. The Green's function matrix is also transformed as $\hat{g}= \hat{U} \hat{g}_q \hat{U}^\dagger$.
To facilitate the solution procedures of the nonlinear Usadel equations we employ $\theta$ parametrization of the Green's functions\cite{Belzig1999},
\begin{equation}
\hat{g}_q= 
\begin{pmatrix}
\cos \theta & \sin \theta \\
\sin \theta & -\cos \theta
\end{pmatrix}.
\end{equation}
Substituting the above matrix into the Usadel equation \eqref{Usadel_general}, we obtain in the S part of the TI surface $x>0$:
\begin{align}
\xi_{s}^2 \pi T_{cs} & \left[ \partial_x^2 \theta_{s} - \frac{q^2 }{2} \sin 2 \theta_s \right]= \\
& =\omega_n \sin{\theta_{s}} - \Delta (x) \cos{\theta_{s}}, \nonumber
\end{align}
and in the F part $x<0$:
\begin{align}
\xi_{f}^2 \pi T_{cs} & \left[ \partial_x^2 \theta_{f} - \frac{q_m^2 }{2} \sin 2 \theta_f \right] =\omega_n \sin{\theta_{f}},
\end{align}
where $q_m = q + 2 h /\alpha$ and $\theta_{s(f)}$ means the value of $\theta$ is the S(F) of the TI surface, respectively. The self-consistency equation for the pair potential reads,
\begin{equation}
\Delta (x) \ln \frac{T_{cs}}{T} = \pi T \sum_{\omega_n>0} \left( \frac{\Delta (x)}{\omega_n} - 2 \sin \theta_s \right).
\end{equation}
To complete the problem formulation we supplement the above equations with the following boundary conditions at $x=0$
\begin{align}
\gamma_B \frac{\partial \theta_f }{\partial x}\Big\vert_{x=0} &= \sin \left( \theta_s - \theta_f \right),\\
\frac{\gamma_B}{\gamma} \frac{\partial \theta_s}{\partial x}\Big\vert_{x=0} &= \sin \left( \theta_s - \theta_f \right),
\end{align}
and at free edges 
\begin{align}
\frac{\partial \theta_f }{\partial x}\Big\vert_{x=-d_f}=0, \quad \frac{\partial \theta_s }{\partial x}\Big\vert_{x=d_s}=0.
\end{align}
In order to calculate the superconducting current we utilize the expression for the supercurrent density
\begin{equation}
\textbf{J}_{s(f)}= \frac{- i \pi \sigma_{s(f)}}{4 e} T \sum_{\omega_n} \tr \left[ \tau_z \hat{g}_{s(f)} \hat{\nabla} \hat{g}_{s(f)} \right].
\end{equation}
Performing the unitary transformation $U$, the current density transforms as follows:
\begin{align}
{j}_y^s (x)=- \frac{\pi \sigma_s q }{2 e} T \sum_{\omega_n} \sin^2 \theta_s, \\
{j}_y^f (x)=- \frac{\pi \sigma_n }{2 e} \left[ q  + \frac{2 h}{\alpha}\right] T  \sum_{\omega_n} \sin^2 \theta_f.
\end{align}
The total supercurrent flowing via the system along the $y$-direction can be calculated by integrated the current density of the total width of the S/F bilayer $d_f+d_s$:
\begin{align}\label{I_total}
I&= \int_{-d_f}^{0} {j}^f_y (x) dx  + \int_{0}^{d_s} {j}^s_y (x) dx .
\end{align}
\begin{figure}[t]
 	\includegraphics[width=\columnwidth,keepaspectratio]{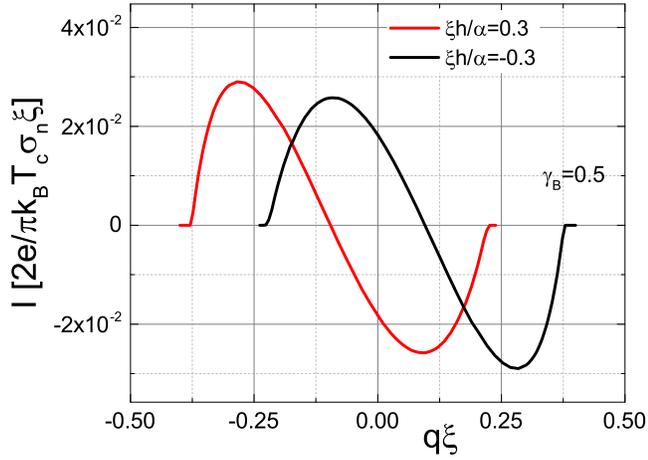}
 	\caption{The normalized supercurrent as a function of $q$ calculated at temperature $T=0.1T_{cs}$  $d_f=1.0 \xi$ and $d_s=1.2 \xi$. The parameters of the S/TI interface: $\gamma=0.5$, $\gamma_B=0.5$.}
 	\label{Curt1}
\end{figure}

\begin{figure}[t]
 	\includegraphics[width=\columnwidth,keepaspectratio]{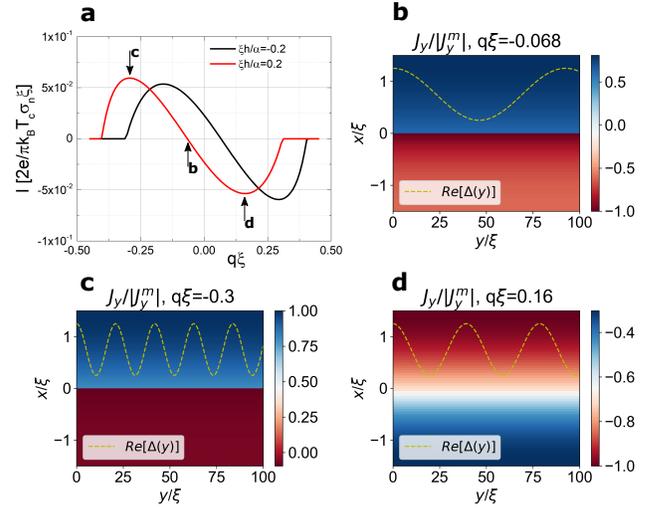}
 	\caption{ (a) The normalized supercurrent as a function of $q$ and  supercurrent density (b-d) calculated at temperature $T=0.1T_{cs}$,  $d_f=1.5 \xi$, $d_s=1.5 \xi$. The parameters of the S/TI interface: $\gamma=0.5$, $\gamma_B=0.5$.}
 	\label{Curt2}
 \end{figure}

In Fig.~\ref{Curt1} the total supercurrent $I$ as a function of the parameter $q$ is shown. We plot the curves for two opposite values of the magnetization strength $h$. Based on the general considerations the function $I(q)$ has a trivial antisymmetric form with respect to $q=0$ in the absence of the exchange field ($h=0$). When $h$ is nonzero the supercurrent loses its purely antisymmetric form, so that the current is finite at $q=0$.
It can be shown that Eq.~(\ref{cond}) is equivalent to the condition
\begin{equation}
I(q_s) = 0 .
\end{equation}
It means that the ground state of the bilayer in the absence of the applied external supercurrent corresponds to the zero total current along the $y$-direction. At the same time the local supercurrent density is not zero. The spatial distribution of the supercurrent at $q=q_s$ (corresponding to the zero current point of the red curve in Fig.~\ref{Curt2} (a)) is demonstrated in Fig.~\ref{Curt2}(b) together with the spatial profile of the real part of the superconducting order parameter. It is seen that in the S/F hybrid  with spatially separated superconductivity and Zeeman field the zero-current helical state is transformed to the kind of a mixed state. It is characterized by the simultaneous presence of the finite pair momentum  and the local supercurrents, which are spatially distributed over the bilayer in such a way to produce zero total current. We call this state  hybrid helical state. 
The above analysis suggests that the bilayer is infinite along the $y$-direction. Therefore we neglect the edge effects. In real setups having a finite length  along the S/F interface the currents should make a U turn at the edges.

\section{Critical current nonreciprocity}\label{I_q}
 Now we investigate the properties of the hybrid helical state under the applied supercurrent.  The maximal supercurrent which is sustained by the system can be extracted from Fig.~{\ref{Curt1}}. Comparing the maximum  absolute values of the positive and negative supercurrents  $I_{c+}$ and $I_{c-}$ , we can recognize that they are distinct in case $h\ne 0$. This is the critical current nonreciprocity $\Delta I_c= I_{c+}-I_{c-}$, which leads to the supercurrent diode effect (SDE). It is only an apparent degeneracy of $q$ with respect to the supercurrent I in Fig.~\ref{Curt1}. The system will choose lower value of $|q|$ since the critical temperature drops as $|q|$ is increased (see Fig.~\ref{fig1}). This situation reminds the well-known problem  of critical current in a superconducting wire, when the relation between current and superfluid velocity is double-valued, but only the state with smallest superfluid velocity is realized \cite{tinkham2004introduction}. We estimate the magnitude of $\Delta I_c$ for the parameters indicated in Fig.~\ref{Curt1} and taking the resistances $\rho_s=0.5 \rho_n= 10 \cdot 10^{-6} Ohm \cdot cm$, the critical temperature $T_{cs}=7 K$, the coherence length $\xi=10 nm$, and $T=0.1 T_{cs}$. For these parameters $\Delta I_c\approx 1.5 \mu A$.

The physics behind the current nonreciprocity can be understood in the following simplified way. In the presence of the exchange field the spin-down states are more energetically favorable. Via the spin-momentum locking it leads to the imbalance between electrons with opposite momenta, what manifests itself as a spontaneous current along the S/F interface. As we have shown, the superconductor produces the counter-propagating current to compensate the current in F. Via this fundamental mechanism of magnetoelectric nature  the exchange field of the ferromagnet influences the phase of the superconducting condensate in the superconducting part of the structure. Now it is natural  that if we adjust the phase gradient $q$ along the $y$-direction via an external source (by applying a supercurrent), it will exert an inverse effect on the effective exchange field. It is clearly seen from the structure of the anomalous Green's function in the F part (Eq.~(\ref{f_T})), where the phase gradient $q$ enters in combination with the exchange field in $k_q=\sqrt{\omega_n + (q+2 h/\alpha)^2}$. If $q$ and $h$ have opposite signs, the spin-polarized electrons, generated by the applied current on the surface of TI,  effectively compensate the suppression of superconducting state by lowering the effective exchange field. Consequently in this case we expect larger possible values of the critical current. However if $q$ and $h$ have the same sign, the generated in TI spin-polarized current flows in the same direction as the equilibrium current, enhancing the effective exchange field $(q+2 h/\alpha)$, which leads to stronger suppression of the superconductivity at the interface. Hence we observe smaller values of the critical supercurrent.
More conveniently the critical current nonreciprocity or the magnitude of the SDE is defined in the dimensionless form as, 
\begin{align}\label{sde}
\delta I = \frac{I_{c+}-I_{c-}}{I_{c+}+I_{c-}}.
\end{align}

\begin{figure}[t]
 	\includegraphics[width=1\columnwidth,keepaspectratio]{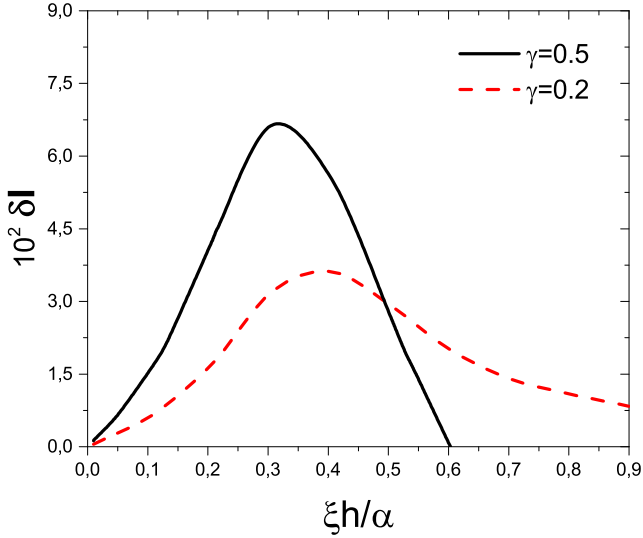}
 	\caption{$\delta I$ as a function of magnetization $h$ calculated for two different $\gamma$ at $T=0.1 T_{cs}$,  $d_f=1.0 \xi$ and $d_s=1.2 \xi$. The interface parameter $\gamma_B=0.5$}
 	\label{dI_1}
 \end{figure}

It is more instructive to discuss SDE by illustrating $\delta I$ dependencies versus various system parameters, including parameters of the proximity effect. In Fig.~\ref{dI_1} we plot $\delta I$ as a function of magnetization $h$ for two different $\gamma$. We see that the dependence of $\delta I$ on h is nonmonotonic.  Such a characteristic behavior is easily explained. At zero exchange field $h$ there is no SDE since the system is not in the helical ground state, but in the conventional superconducting state with a homogeneous phase. As the exchange field increases the SDE also rises but eventually starts to drop due to suppression of the superconductivity by the field $h$.

\begin{figure}[t]
 	\includegraphics[width=1\columnwidth,keepaspectratio]{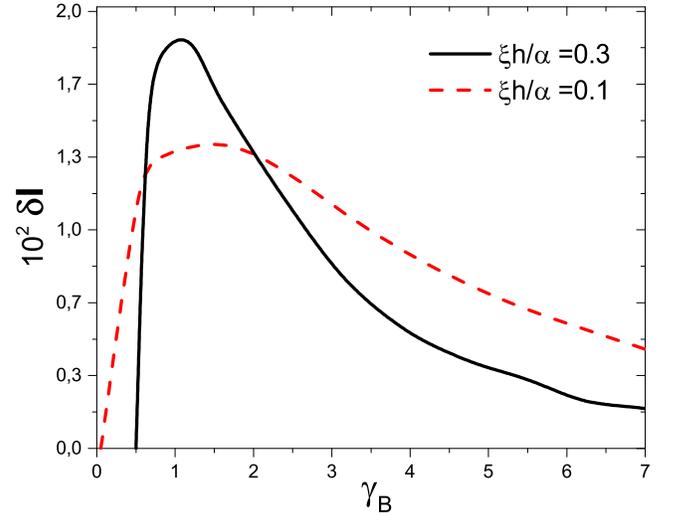}
 	\caption{$\delta I$ as a function of transparency parameter $\gamma_B$ calculated at two different $h$ for  $d_f=1.0 \xi$ and $d_s=1.2 \xi$. The temperature is taken as $T=0.4 T_{cs}$ and $\gamma =0.5$}
 	\label{dI_2}
 \end{figure}

There are possibilities to design the superconducting diode not only by tuning the magnetization $h$, which can be quite challenging in practice, but by adjusting other parameters such as $\gamma_B$. In Fig.~\ref{dI_2} $\delta I (\gamma_B)$ dependencies are shown. Here we observe a nonmonotonic dependence of the SDE on the interface transparency $\gamma_B$. The decay of the SDE at large $\gamma_B$ is physically clear because increase of the interface transparency reduces the mutual proximity influence of the spatially separated exchange field and superconductivity. On the contrary at relatively small values of $\gamma_B$ superconductivity can be substantially suppressed (red dashed line) or even completely vanish (black solid line).

\begin{figure}[t]
 	\includegraphics[width=1\columnwidth,keepaspectratio]{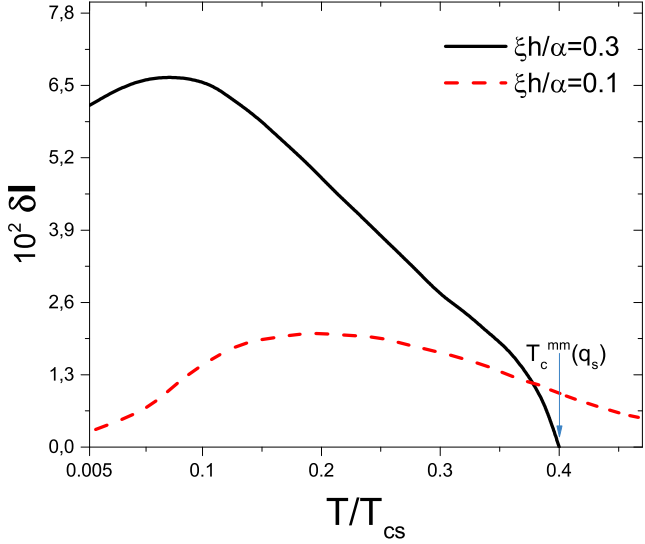}
 	\caption{$\delta I$ as a function of temperature $T$ calculated for two different $\gamma$. The curves were calculated for $d_f=1.0 \xi$, $d_s=1.2 \xi$, $\gamma=0.5$ and $\gamma_B=0.5$. Here $T_c^{mm}$ is the transition temperature obtained via the multimode approach (see Appendix).}
 	\label{dI_3}
 \end{figure}

We also illustrate the nonreciprocity of the current $\delta I$ as a function of the system temperature $T$ (Fig.~\ref{dI_3}). It is interesting that the dependence $\delta I(T)$ is nonmonotonic. Similar type of behavior has been also found in the ballistic Rashba superconductors\cite{Ilic_arxiv}. The critical temperature indicated in the plot is in the correspondence with $T_c$ calculated by the multimode approach.

\begin{figure}[t]
 	\includegraphics[width=1\columnwidth,keepaspectratio]{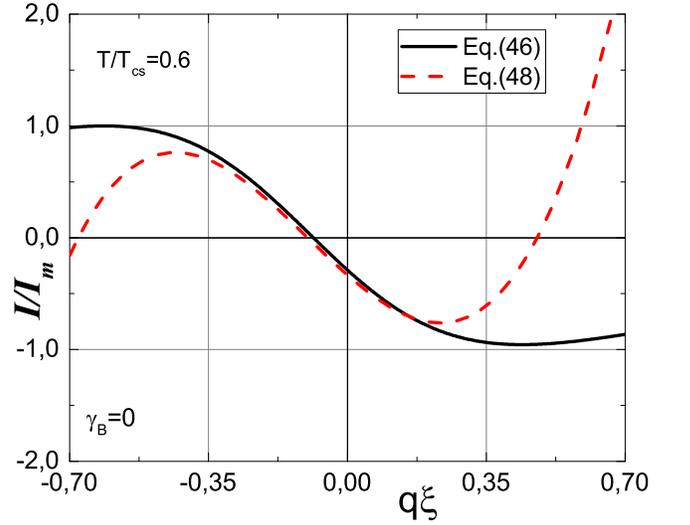}
 	\caption{Analytical results for $I(q)$ calculated according to Eq.~\eqref{I_a_1} - black solid line and Eq.~\eqref{I_a_2} - red dashed line. Here $I_m$ corresponds to the maximum value of the current calculated from Eq.~\eqref{I_a_1}. The parameters of the S/TI interface: $\gamma=1$ and $\gamma_B=0$. The rest of the parameters are: $d_s=0.5 \xi$, $d_f=0.5 \xi$ and $\xi h/\alpha =0.1$ }
 	\label{Analyt}
 \end{figure}

 In the framework of the linear Usadel equations under the assumption $d_s \ll \xi_s$,  we can easily find the total supercurrent integrating contributions from the S part and F part of the junction.
Substituting the solutions Eqs.~\eqref{thins_sol}-\eqref{F_sol_an} into the current formula and performing integration, one obtains the averaged supercurrent in the Cooper limit ($\Delta= \const$),
\begin{align}\label{I_a_1}
    I &=\frac{\pi \Delta^2 \sigma_s T \xi^3}{2 e} \sum_{\omega_n} \frac{I_f\left(q \xi  +H \right) + I_s q \xi }{(\omega_n/ \pi T_{cs} + (q \xi)^2)^2},\\
    I_s&= d_s - 2 P \frac{A_{qs}}{k_{qs}^2} +P^2 \left(\frac{d_s}{2 \cosh^2{k_{qs} d_s}} +  \frac{A_{qs}}{2 k_{qs}^2}\right), \nonumber \\
    I_f&= \gamma \left(\frac{1- P}{\gamma_B + A_{qT}}\right)^2 \left( \frac{d_f}{2 k_q^2 \xi^2 \sinh^2{k_q d_f}} + \frac{\coth{k_q d_f}}{2 k_q^3 \xi^2}\right), \nonumber
\end{align}
where $P=W^{q}(\omega_n)/(W^{q}(\omega_n)+A_{qs})$. In the limit of small TI layer widths $d_f$, perfect interface transparency $\gamma_B=0$ and strong proximity effect $\gamma=1$, we can write  $W^{q}(\omega_n)$ ( Eq.~\eqref{W_om}) in a more simplified way,
\begin{align}
W^{q}(\omega_n) \approx \frac{1}{A_{qT}} \approx \frac{d_f}{\xi} \left( \frac{\omega_n}{ \xi^2 \pi T_{cs}} + \left( q \xi  + H \right)^2  \right).
\end{align}
Assuming $q \xi \ll 1$  and keeping the terms up to the third order we can derive analytical expression for the total supercurrent summing both S and TI layer contributions. The supercurrent is then,
\begin{align}\label{I_a_2}
    I&= -\frac{ \pi \Delta^2 \sigma T}{2 e} \left(a_0 + a_1 q \xi + a_2 (q \xi)^2 + a_3 (q \xi)^3 \right), \\
    a_0&=\frac{d_f d_s^2}{ \xi^3}\left(\frac{H \xi^2}{d^2} \sum \frac{\left(\pi T_{cs}\right)^2}{\omega_n^2} - \frac{2 H^3 d_f \xi }{d^2} \sum \frac{\left(\pi T_{cs}\right)^3}{\omega_n^3}\right),\nonumber \\
    a_1&= \frac{d_s^2}{d \xi} \sum \frac{\left(\pi T_{cs}\right)^2}{\omega_n^2} - \frac{2 H^2}{d^3 \xi} (3 d_f^2 d_s^2 + d_f d_s^3) \sum \frac{\left(\pi T_{cs}\right)^3}{\omega_n^3}, \nonumber \\
    a_2&= \frac{2 H}{d^3 \xi} (d_f d_s^3 - 3 d_f^2 d_s^2) \sum \frac{\left(\pi T_{cs}\right)^3}{\omega_n^3}, \nonumber \\
    a_3&= - \frac{2 d_s^2}{d^3 \xi} (2 d_f^2 + d_f d_s + d_s^2) \sum \frac{\left(\pi T_{cs}\right)^3}{\omega_n^3}. \nonumber
\end{align}
Here, we have denoted  $d=(d_s + d_f)$. In Fig.~\ref{Analyt} we demonstrate  the analytical calculations in the Cooper limit. The solid line corresponds to Eq.~\eqref{I_a_1}, which is valid for arbitrary TI layer width $d_f$, interface parameters $\gamma$ and $\gamma_B$. The red dashed line represents Eq.~\eqref{I_a_2}. From the figure we can say that Eq.~\eqref{I_a_2} is in a relatively good agreement with the more general formula at small values of $q$. However, it fails at larger values of $q$. In order to describe larger $q$ successfully, one must take into account the terms of the next orders.

From Eq.\eqref{I_a_2} one can derive analytically $I_{c+}$ and $I_{c-}$ by applying the  extremum condition to $I(q)$,
\begin{align}
    \frac{d I}{d q}= a_1 + 2 a_2 (q \xi) + 3 a_3 (q \xi)^2 =0.
    \label{extremum_analytical}
\end{align}
Solution of Eq.~(\ref{extremum_analytical}) yields,
\begin{align}
    I_{c\pm} = a_0 - \frac{1}{3} q_{\pm}^2 \left( a_2 \mp 2\sqrt{a_2^2 + a_1 |a_3| }\right).
\end{align}
The complete expression for the SDE $\delta I$ is rather cumbersome to display here. Instead we can find relatively simple result in the limit of $H d_f/\xi \ll 1 $. In this case we obtain that
\begin{align}\label{dI_a}
    \delta I \approx \frac{1}{2}  \frac{\sqrt{7 \zeta (2) \zeta (3) }}{\left(T/T_{cs}\right)^{5/2}} \frac{H d_f}{d_s} \approx 1.86 \frac{1}{\left(T/T_{cs}\right)^{5/2}} \frac{H d_f}{d_s}.
\end{align}
This result demonstrates that the SDE is controlled by the product $(H d_f /d_s)$. Moreover it can be noticed that Eq.~\eqref{dI_a} reveals the temperature dependence, showing characteristic scaling behavior of the SDE as a function of temperature. Please note that Eq.~(\ref{dI_a}) is not valid at $T \to 0$, where our linearized Usadel theory does not work.

\section{Discussion and Conclusion}\label{Fin}
 We have examined the  characteristic features of the superconducting helical state in the S/F/TI hybrid structure with an in-plane exchange field perpendicular to the interface. It has been found that the ground state of the system is characterized by the superconducting order parameter modulated with finite momentum $q_s$. At the same time due to the spatial separation of the superconductivity and ferromagnetism in the hybrid structure this state is accompanied by the non-zero current distribution. The currents flow along the S/F interface and are distributed over the whole structure. The current distribution corresponds to zero net value of the current along the S/F interface. We have found that this hybrid helical state is responsible for substantial nonreciprocity of the critical current in the system due to strong spin-orbit coupling on the surface of TI. Direct manifestation of the nonreciprocity is the superconducting diode effect. Finally, we have derived important analytical results, revealing controlling parameters and temperature dependence of the SDE.
 
 The nonlinear self-consistent Usadel equations employed in this study is a relatively simple but powerful method for treating such systems. Since we have considered the diffusive limit in our model, as a further step it is important to analyze the problem in the ballistic limit and make corresponding comparisons between the two models.
\begin{acknowledgments}
The work was supported by RSF project No. 18-72-10135. I.V.B. and T.K. acknowledge the financial support by the Foundation for the Advancement of Theoretical Physics and Mathematics “BASIS”. T.K. and A.S.V. acknowledge support from the Mirror Laboratories Project and the Basic Research Program of HSE University.
\end{acknowledgments}

\appendix*
\section{Multimode approach}
\begin{figure*}[t]
	\centering
	\includegraphics[width=2\columnwidth,keepaspectratio] {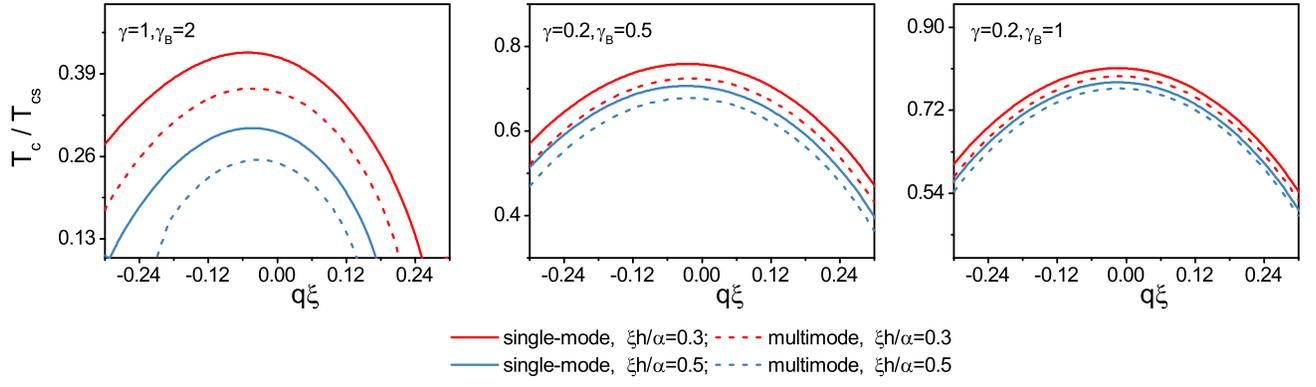}
	\caption
	{
	$T_c (q)$ curves calculated with single (solid lines) and multimode approaches (dashed lines) at different interface parameters $\gamma$ and $\gamma_B$. The multimode curves are calculated at $N=100$.
	}
	\label{Tc_multi}
\end{figure*}
Here we present the multimode method to solve the critical temperature problem \cite{Fominov2002,Karabassov2019}. The single-mode approach takes into account only one real root provided by Eq.~\eqref{last1}. In order to introduce exact solving method for the problem under consideration one also takes imaginary roots $\Omega_m$ ($m>0$) into account apart from the real root. In general the number of roots is infinite.

In the framework of the multimode method the solution of Eqs.~\eqref{Delta+}-\eqref{finUsadelS} is found in the form,
\begin{align}
F_s^+(x,\omega_n) = & f_0(\omega_n)\cos\left(\Omega_0 \frac{x-d_s}{\xi_s} \right) \nonumber \\
&+ \sum_{m=1}^{\infty}f_m(\omega_n)\frac{\cosh\left(\Omega_m \frac{x-d_s}{\xi_s} \right)}
{\cosh\left(\Omega_m \frac{d_s}{\xi_s}\right)}, \label{Phi}
\end{align}
\begin{align}
\Delta(x) = & \delta_0 \cos\left(\Omega_0\frac{x-d_s}{\xi_s}\right) \nonumber \\ 
&+ \sum_{m=1}^{\infty} \delta_m \frac{\cosh\left(\Omega_m\frac{x-d_s}{\xi_s}\right)}
{\cosh\left(\Omega_m\frac{d_s}{\xi_s}\right)}. \label{Delta}
\end{align}
The solution presented by the multimode approach automatically satisfies boundary condition at $x=d_s$. After the substitutions into the Usadel equation in the S part \eqref{finUsadelS} $f(\omega_n)$ can be expressed as,
\begin{align*}\label{f_om_0}
f_0(\omega_n)=\frac{2 \delta_0}{\omega_n + \Omega_0^2 \pi T_{cs} + q^2 \xi_s^2 \pi T_{cs}}
\end{align*}
\begin{align}
f_m(\omega_n)=\frac{2 \delta_m}{\omega_n - \Omega_m^2 \pi T_{cs} + q^2 \xi_s^2 \pi T_{cs}},      \quad m=1,2,...
\end{align}
Then the self-consistency equation \eqref{Delta+} takes form
\begin{equation*}
\ln \frac{T_{cs}}{T_c} = \psi \left ( \frac{1}{2} + \frac{\Omega_0^2 + q^2 \xi_s^2}{2}\frac{T_{cs}}{T_c} \right ) - \psi \left ( \frac{1}{2} \right ),
\end{equation*}
\begin{equation}\label{dig}
\ln \frac{T_{cs}}{T_c} = \psi \left ( \frac{1}{2} - \frac{\Omega_m^2 - q^2 \xi_s^2}{2}\frac{T_{cs}}{T_c} \right ) - \psi \left ( \frac{1}{2} \right ), \quad m=1,2,...
\end{equation}
According to properties of digamma function and Eqs.~\eqref{dig} it follows that the parameters $\Omega$ belong to the following intervals:
\begin{equation*}
0<\Omega_0^2<\frac{1}{2 \gamma_E},
\end{equation*}
\begin{equation}
\frac{T_c}{T_{cs}} \left(2m-1\right)<\Omega_m^2< \frac{T_c}{T_{cs}} \left(2m+1\right), \quad m=1,2,...,
\end{equation}
where $\gamma_E\approx1.78$ is Euler's constant . Boundary condition \eqref{B1a} at $x=0$ provides the equation for the amplitudes $\delta$
\begin{multline}\label{deltaeq}
\delta_0 \frac{W^{q}(\omega_n)\cos\left(\Omega_0 \frac{d_s}{\xi_s}\right)- \Omega_0 \sin\left(\Omega_0 \frac{d_s}{\xi_s}\right)}{\omega_n + \Omega_0^2\pi T_{cs} + q^2 \xi_s^2 \pi T_{cs}}\\+ \sum_{m=1}^{\infty} \delta_m \frac{W^{q}(\omega_n)+ \Omega_m \tanh\left(\Omega_m \frac{d_s}{\xi_s}\right)}{\omega_n - \Omega_m^2\pi T_{cs} + q^2 \xi_s^2 \pi T_{cs}}=0.
\end{multline}
The critical temperature $T_c$ is calculated by Eqs.\eqref{dig} and Eq.\eqref{deltaeq}. In order to solve the problem numerically one takes finite number of roots $\Omega$ with $m=0,1,2...,M$, also taking into account Matsubara frequencies $\omega_n$ up to the $N$th frequency: $n=0,1,2...,N$. Hence the matrix equation has the following form: $K_{nm}\delta_m=0$ with matrix $\operatorname{K}$
\begin{equation*}
K_{n0}=\frac{W^{q}(\omega_n)\cos\left(\Omega_0 \frac{d_s}{\xi_s}\right)- \Omega_0 \sin\left(\Omega_0 \frac{d_s}{\xi_s}\right)}{\omega_n/\pi T_{cs} + \Omega_0^2 + q^2 \xi_s^2 \pi T_{cs}},
\end{equation*}

\begin{equation}
K_{nm}=\frac{W^{q}(\omega_n)+ \Omega_m \tanh\left(\Omega_m \frac{d_s}{\xi_s}\right)}{\omega_n/\pi T_{cs} - \Omega_m^2 + q^2 \xi_s^2 \pi T_{cs}},
\end{equation}
\begin{align*}
n=0,1,...,N, \quad m=1,2,...,M.
\end{align*}
We take $M=N$ , then the condition that equation \eqref{deltaeq} has nontrivial solution takes the form
\begin{equation}\label{detk}
\det K=0.
\end{equation}
Now we compare the results obtained by single-mode and multimode approaches by calculating $T_c(q)$ dependencies. From Fig.~\ref{Tc_multi} we can see that the two methods produce quantitative differences at relatively large $\gamma$ and small $\gamma_B$. Nevertheless,  the results are not affected qualitatively. As we increase the interface resistance $\gamma_B$ or reduce $\gamma$ the quantitative discrepancy vanishes. 

%
\bibliography{diode.bib}
\end{document}